\documentclass[a4paper,10pt]{article}
\usepackage[utf8]{inputenc}
\usepackage{amsmath}
\usepackage{amsfonts}
\usepackage{graphicx}
\usepackage{subcaption}
\usepackage{bbold}
\usepackage{geometry}
\usepackage{color}

\title{Black holes and black strings in the Einstein $SU(N)$-non-linear sigma model}
\author{Carla Henr\'iquez-B\'aez$^{1}$, Marcela Lagos$^{2}$, Aldo Vera$^{2}$ \\
$^{1}$\textit{Departamento de Matem\'atica y F\'isica Aplicadas, Universidad Cat\'olica de la Sant\'isima Concepci\'on,}\\ \textit{Concepci\'on, Chile}\\
$^{2}$\textit{Instituto de Ciencias F\'isicas y Matem\'aticas, Universidad
Austral de Chile,}\\\textit{ Casilla 567, Valdivia, Chile}\\
{\small carlahenriquez@udec.cl, marcela.lagos@uach.cl,
aldo.vera@uach.cl}}
\begin{document}
\maketitle
\begin{abstract}
We construct analytical solutions describing black holes and black strings in the Einstein $SU(N)$-non-linear sigma model in $(3+1)$ dimensions.
This construction is carried out using the maximal embbeding ansatz of $SU(2)$ together with the Euler parameterization of the $SU(N)$ group,
in such a way that the non-linear sigma model equations are automatically satisfied for arbitrary values of the flavor number $N$ while the
Einstein equations can be solved analytically. 
In particular, we construct black holes with spherical and flat horizons as well as black strings that present the geometry of a three-dimensional
charged Bañados-Teitelboim-Zanelli black hole on the transverse section of the string. These configurations are not
trivial embeddings of $SU(2)$ into $SU(N)$, which allow us to explicitly show the role that the flavor number plays on the geometry and thermodynamics
of the black holes and black strings. Finally, we perform a thermal comparison between these configurations.
\end{abstract}

\newpage 

\tableofcontents


\section{Introduction}


The non-linear sigma model (NLSM) is one of the most important effective field theories, both from the theoretical point of view as well 
as phenomenologically, whose applications range from 
quantum field theory to statistical mechanics and string theory \cite{manton}.
One of the main applications of the NLSM is the description of the low-energy dynamics of pions \cite{Nair},
which is performed considering as internal symmetry the $SU(2)$ Lie group, that is, the two flavors case.

Now, although the predictions provided by the model are in good agreement with the experimental results,
obtaining such predictions is not a simple task since the field equations that emerge from the NLSM are in general very complicated. In fact, these are a set of $(N^2-1)$ coupled non-linear differential equations 
-being $N$ the flavor number encoded in the $SU(N)$ group- which explains that most of the known solutions have been constructed numerically. Some relevant results in this approach are found in Refs. \cite{Droz}, \cite{LM}, \cite{v1} and \cite{v2}. Of course, when the theory is coupled to general relativity\footnote{When a suitable interaction potential is considered some interesting analytic solutions can be constructed; see \cite{Anabalon}.} or the Maxwell theory to describe more complex physical objects and processes, the field equations become even more intrincate. However, in Refs. \cite{Canfora} and \cite{CanforaMaeda}, suitable ans\"atze have been introduced that go beyond
spherical symmetry (collectively called generalized hedgehog ansatz), 
allowing to find a good number of exact solutions not only in the NLSM, but also in the Skyrme model,
the generalized Skyrme model and the Yang-Mills-Higgs theory.
These new set of solutions describe boson stars \cite{Cavity}, black holes \cite{CanforaMaeda}, \cite{ACGO}, \cite{Rojas},\footnote{See \cite{Eiroa} for the analysis of the lensing produced by these Skyrme black holes.} black strings \cite{ACLV}, gravitating solitons
\cite{Eloy1}, \cite{Eloy2}, topological solitons at finite volume \cite{Alvarez}, and even crystalline structures of topological solitons \cite{crystal1}, \cite{crystal2} (see also \cite{GO}, \cite{Canfora:2021rdw}, \cite{Canfora:2021dlk}, \cite{Canfora:2021nca}, \cite{tube}, \cite{Dotti} and references therein).

Although it is quite remarkable that such equations can be solved analytically, it is important to note that most of these solutions have been found in the two flavors case, where three degrees of freedom are present. Building solutions with $N>2$ is a more complicated task, as can be seen in Refs. \cite{Eloy2}, \cite{Bala1} and \cite{Bala2}. The main goal of this work is to construct analytical solutions of the Einstein-NLSM theory for arbitrary values of $N$. In particular, we will construct solutions that describe black holes and black strings by combining the generalized hedgehog ansatz with the Euler angle parameterization through the maximal embbeding of $SU(2)$. This maximal embedding, introduced in Refs. \cite{euler1}, \cite{euler2} and \cite{euler3}, allows us to construct in a direct way, genuine $SU(N)$ fields that are not trivial embeddings of $SU(2)$ into $SU(N)$. The power of this formalism can be seen in Refs. \cite{SU(N)1}, \cite{SU(N)2}, \cite{SU(N)3} and \cite{Gomberoff}, where nuclear pasta states in the $SU(N)$-Skyrme model
and black holes in the Einstein $SU(N)$-Yang-Mills theory have been constructed.
We will see that the pionic black objects constructed here possess a non-trivial thermodynamics and intriguing geometrical properties. In this way, we will generalize to the $SU(N)$ case the solutions constructed in Refs. \cite{CanforaMaeda}, \cite{ACGO} and \cite{ACLV} for the NLSM, explicitly showing the role that the flavor number plays both in the thermodynamics and in the geometry of these configurations. 

In particular, the black string solution in Ref. \cite{ACLV}, and its generalization to the $SU(N)$ group constructed here, is of great interest.
As it is well known, black hole solutions in general relativity in four dimensions have very well defined properties including restrictions of the topology of the horizons. But, in higher dimensions, extended objects as black string and $p$-branes arise naturally and exhibit properties that differ from those of the black holes \cite{HorowitzBook}, \cite{Lemos}, \cite{Vanzo}, \cite{GibbonsWiltshire}, \cite{Horowitz}, \cite{Myers}, \cite{Emparan1}. Between these novel properties, it was demonstrated that these configurations present a long wavelength instability, producing in its final state a naked singularity \cite{GregoryLaflamme1}, \cite{GregoryLaflamme2},
\cite{Lehner}, \cite{Emparan2} (see also \cite{GL1}, \cite{GL2}), leading to an explicit violation of cosmic censorship.

However, the presence of additional fields and/or a cosmological constant, may allows to construct solutions with a variety of topologies of the event horizon beyond spherical symmetry (see, for instance, \cite{BS1}, \cite{BS2}, \cite{BS3}, \cite{Cisterna:2021ckn}, \cite{Corral}) even in four dimensions. This is the case of the black string presented here. Interestingly, the presence of additional fields living on the extended directions of the string has shown to be very useful to stabilize this type of solutions \cite{BS4}. In the present work we also discuss some relevant issues about the stability of our solutions focusing on the role played by the parameter $N$.

The paper is organized as follows: In Sec. 2 we give a brief review of the Einstein-NLSM theory and present our general ansatz for the $SU(N)$ matter field. 
In Sec. 3 we construct analytical solutions describing black holes and black strings and discuss its main physical properties.
In Sec. 4 we compare the solutions through a thermal analysis. In the last section we draw some conclusions.


\section{Preliminaries}


\subsection{The Einstein $SU(N)$-non-linear sigma model and its field equations}

The Einstein $SU(N)$-NLSM in $(3+1)$ dimensions is described by the action 
\begin{equation}
I[g,U]=\int d^{4}x\sqrt{-g}\left( \frac{R}{2\kappa }+\frac{K}{4}%
\mathrm{Tr}[L^{\mu }L_{\mu }]\right) ,  \label{I}
\end{equation}%
where $R$ is the Ricci scalar and $L_{\mu }$ are the Maurer-Cartan
form components
\begin{equation} \label{L}
L_{\mu }=U^{-1}\partial _{\mu }U = L_\mu^i t_i \ ,
\end{equation}
for $U(x) \in SU(N)$, being $N$ the flavor number and $t_i$ the generators of the $SU(N)$ Lie group, with $i=1,..., (N^2-1)$.
Here $\kappa$ is the gravitational constant and $K$ is a positive coupling fixed by experimental data. In our convention $c=\hbar=1$, Greek
indices run over the four dimensional space-time with mostly plus signature,\footnote{In this paper, we follow the standard convention that the Riemann curvature
tensor, the Ricci tensor, and the Ricci scalar are given by 
\begin{gather*}
R^{\alpha}_{\phantom{\alpha} \beta \mu \nu} = \Gamma^{\alpha}_{%
\phantom{\alpha} \beta \nu , \mu} -\Gamma^{\alpha}_{\phantom{\alpha} \beta
\mu, \nu} +\Gamma^{\alpha}_{\phantom{\alpha} \rho \mu} \Gamma^{\rho}_{%
\phantom{\rho} \beta \nu} -\Gamma^{\alpha}_{\phantom{\alpha} \rho \nu}
\Gamma^{\rho}_{\phantom{\rho} \beta \mu} \ , \\
R_{\mu \nu} = R^{\alpha}_{\phantom{\alpha} \mu \alpha \nu} \ , \qquad 
R = g^{\mu \nu} R_{\mu \nu} \ .
\end{gather*}
} while Latin indices are reserved for those of the
internal space.

The field equations of the model, obtained varying the action in Eq. \eqref{I} w.r.t the fundamental fields $U$ and $g_{\mu\nu}$, are 
\begin{gather} \label{Eq1}
\nabla ^{\mu }L_{\mu }=0 \ , \\ 
G_{\mu \nu }=\kappa T_{\mu \nu }\ , \label{Eq2}
\end{gather}%
where $G_{\mu \nu }$ is the Einstein tensor, $\nabla_\mu$ denotes the Levi-Civita covariant derivative, and $T_{\mu \nu }$ is the
energy-momentum tensor of the NLSM, given by 
\begin{equation}  \label{T}
T_{\mu \nu }=-\frac{K}{2}\mathrm{Tr}\left[ L_{\mu }L_{\nu }-\frac{1}{2}%
g_{\mu \nu }L^{\alpha }L_{\alpha }\right] .
\end{equation}%
Note that in Eq. \eqref{Eq1} are $(N^2-1)$ non-linear coupled second order differential equations.

\subsection{General ansatz for the matter field}

The main goal of this paper is to construct analytical solutions with internal symmetry group genuinely $SU(N)$, that is, configurations with $U(x)$ in a subgroup of $SU(N)$ that is a non trivial embbeding of $SU(2)$ into $SU(N)$.
In order to do that, we will use the so-called maximal embedding \cite{euler1}, \cite{euler2}, \cite{euler3}, which gives rise to an irreducible representation
of $SU(2)$ of spin $j = (N - 1) / 2$.

For the matter field $U$ we will use the Euler angle representation, that is
\begin{equation} \label{U}
 U \ = \ e^{F_1(x^\mu) \cdot T_3} e^{F_2(x^\mu) \cdot T_2} e^{F_3(x^\mu) \cdot T_3} \ , 
\end{equation}
where $\{T_{1},T_{2},T_{3}\}$ are three matrices of a given
representation of the Lie algebra $su(N)$, which will be chosen in order to
satisfy the following relations 
\begin{equation*}
\lbrack T_{j},T_{k}]=\epsilon _{jkm}T_{m}\ ,\qquad \text{Tr}(T_{j}T_{k})=-%
\frac{N(N^{2}-1)}{12}\delta _{jk}\ .
\end{equation*}
The above matrices $T_i$ define a three dimensional subalgebra of $su(N)$, and they are given explicitly as 
\begin{align}
T_1&=-\frac{i}{2}\sum_{j=2}^{N} \sqrt{(j-1)(N-j+1)}(E_{j-1,j}+E_{j,j-1}) \ ,
\label{T1} \\
T_2&=\frac{1}{2}\sum_{j=2}^{N} \sqrt{(j-1)(N-j+1)}(E_{j-1,j}+E_{j,j-1}) \ ,
\label{T2} \\
T_3&=i\sum_{j=1}^{N} (\frac{N+1}{2}-j)E_{j,j} \ ,  \label{T3}
\end{align}
with 
\begin{equation*}
(E_{i,j})_{mn}=\delta_{im}\delta_{jn}\ ,
\end{equation*}
being $\delta_{ij}$ the Kronecker delta.\footnote{Note that the $T_i$ matrices are antihermitian. But, one can easily recover an hermitian set by multiplying the matrices by $i$. Of course, in order to obtain the same solutions presented below we need to multiply the $F_i$ function also by $i$.} The complete mathematical formulation of the above construction can be found
in Refs. \cite{euler1}, \cite{euler2} and \cite{euler3}
(see also \cite{SU(N)1}, \cite{SU(N)2}, \cite{SU(N)3} and \cite{Gomberoff} for recent applications in physics).

It is worth to emphasize that the above generators form an irreducible representation of the $SU(2)$ group.
In fact, one can note that 
\begin{equation}
 (\vec{T})^2= \rho(N) \mathbb{1} \ , \qquad \rho(N)=-\frac{N^2-1}{4} \ ,
\end{equation}
with $\mathbb{1}$ the $N\times N$ identity matrix. Here the flavor number appears explicitly and, therefore, the solutions that we will construct, 
using the above ansatz, are non trivial embeddings of $SU(2)$ into $SU(N)$.\footnote{This is not the case for any embedding in $SU(N)$.
For instance, in the $SU(3)$ case, one may take one half of the first three Gell-Mann matrices as generators of $SU(2)$, which form a spin $1/2$ representation of $SU(2)$, but this is not irreducible because its three $3\times3$ matrices have zeros everywhere except for their $2\times2$ first blocks.}
In other words, here we will consider ansatz (for all $N$) constructed from just three matrices
(see Eq. \eqref{U}), but which can not be reduced to simply elements of $SU(2)$ times identity matrices. This assumption is fundamental, because it leads to a system of equations simple enough to be solved analytically, but at the same time suficiently rich, such that it is possible to explicitly see the role that $N$ plays in the physics of these configurations.  

On the other hand, in principle the functions $F_i $ in Eq. \eqref{U} can depend on all the coordinates,
but they will be chosen appropriately in order to solve Eqs. \eqref{Eq1} and \eqref{Eq2} analytically. In particular, Eq. \eqref{Eq1} will be identically satisfied as we will see below. The ansatz used in this work for the $U$ field only depends on two spatial coordinates, namely $\theta$ and $\phi$, whose ranges are
\begin{equation} \label{ranges}
 0 \leq \theta < \pi \ , \qquad 0 \leq \phi < 2\pi \ . 
\end{equation}
A linear dependence on the angular coordinates of the functions $F_i$ in Eq. \eqref{U} is one
of the key points that allows obtaining analytical solutions of the Einstein $SU(N)$-NLSM system for arbitrary values of $N$.

Finally, for what follows, it is useful to define the following positive quantity 
\begin{equation} \label{aN}
 a_{N} = \frac{N(N^2-1)}{6} \ ,
\end{equation}
which will appear repeatedly in both, the black hole and black string solutions. 


\section{Analytic black holes and black strings}


In this section we will present black hole and black string configurations that are analytical solutions
of the Einstein $SU(N)$-NLSM for arbitrary values of $N$. As starting point we will use the ansatz in Eq. \eqref{U} for the matter field.

\subsection{Spherical black hole}

Let us consider a spherically symmetric space-time characterized by the metric 
\begin{equation}      \label{metric1}
 ds^2=-f(r) dt^2 +\frac{1}{f(r)} dr^2 + r^2 d\theta^2 +r^2 \sin^2{\theta}d\phi^2 \ ,  
\end{equation}
together with a matter field $U$ in Eq. \eqref{U} with the following particular form for the $F_i$ functions
\begin{equation} \label{F1}
 F_1(x^\mu)=-p \phi \ , \quad F_2(x^\mu)=2q \theta \ , \quad F_3(x^\mu)= p \phi \ ,
 \end{equation}
where $p$ and $q$ are real constants.

One can check that, replacing the ansatz in Eqs. \eqref{U}, \eqref{metric1} and \eqref{F1}, the NLSM field equations in Eq. \eqref{Eq1} are satisfied, while the Einstein equations can be solved analytically, obtaining
\begin{equation} \label{f1}
 f(r)= 1-  K\kappa a_{N} -\frac{2m}{r}-\frac{\Lambda}{3}r^2 \ .
\end{equation}
Here $m$ is an integration constant (related to the mass) and $a_N$ has been defined in Eq. \eqref{aN}.
In this case (namely, for the solution with spherical symmetry in our family of solutions), the Einstein equations also lead to the constraint
$p=q=1$ in Eq. \eqref{F1}, therefore, this configuration does not have further parameters on top of the mass that can be considered as hair. This is different from the case of the flat black hole and the black string, as we will see in the next subsections. 
Clearly, the above configuration represents a spherically symmetric black hole supported by pionic matter,
and it is the generalization to the $SU(N)$ case of the black hole found in Refs. \cite{CanforaMaeda} (without the Skyrme term) and \cite{Gibbons}. 
Also, the above solution is asymptotically the Anti-de Sitter version of the Barriola–Vilenkin space-time \cite{BV}, and it generalizes the Schwarzschild
Anti-de Sitter space-time, which is recovered by setting $K=0$.

As with the black hole in Refs. \cite{CanforaMaeda} and \cite{Gibbons}, the generalization to $SU(N)$ also has an angular defect, and it depends directly on the value of $N$, 
which can be seen in Eq. \eqref{f1}. As it is well known, the presence of an angular defect affects the calculation of the mass and the thermodynamic quantities
of the system, therefore, the extended thermodynamic formalism must be applied (see next section). 

In these coordinates,\footnote{Later, to compute the thermodynamics, we will introduce a new coordinate system to remove the angular defect from  Eq. \eqref{f1}.} the event horizon of the black hole is localized at
\begin{equation} \label{rmas1}
\tilde{r}_{+} =\frac{-\Lambda+K \kappa a_{N} \Lambda-\left(3m\Lambda^2+\sqrt{\Lambda^3\left(\left(K \kappa a_{N}-1\right)^3+9m^2\Lambda\right)}\right)^{2/3}}{\Lambda\left(3m\Lambda^2+\sqrt{\Lambda^3\left(\left(K \kappa a_{N}-1\right)^3+9m^2\Lambda\right)}\right)^{1/3}} \ , 
\end{equation}
where we can see that, to have a real square root in the previous expression, the integration constant $m$ must satisfies the following relation,
\begin{equation*}
m\geq \frac{\left(K\kappa a_{N}-1\right)^{3/2}}{3 \sqrt{-\Lambda}} \ .
\end{equation*}
The above relation determines a minimum radius of the event horizon, namely
\begin{equation}   \label{rminsph}
 \tilde{r}_{\text{min}} = \frac{2\sqrt{K \kappa a_{N} -1 }}{\sqrt{-\Lambda}}  \ . 
\end{equation}

\subsection{Flat black hole}

A flat black hole can be constructed considering a static space-time with a flat base manifold given by the metric
\begin{equation}      \label{metric2}
 ds^2=-f(r) dt^2 +\frac{1}{f(r)} dr^2 + r^2 d\theta^2 + c_0^2 r^2 d\phi^2 \ ,  
\end{equation}
with $c_0$ a constant to be fixed. For the matter field in Eq. \eqref{U} we consider the following ansatz for the $F_i$ functions
\begin{equation} \label{F2}
 F_1(x^\mu)=0 \ , \quad F_2(x^\mu)=q\theta \ , \quad F_3(x^\mu)=p\phi \ .
 \end{equation}
Putting the ansatz in Eqs. \eqref{U}, \eqref{metric2} and \eqref{F2}, into Eqs. \eqref{Eq1} and \eqref{Eq2}, the complete set of field equations of 
the Einstein $SU(N)$-NLSM are solved by
 \begin{equation} \label{f2}
 f(r)= -\frac{ q^2 K\kappa  a_{N} }{4} -\frac{m}{r} -\frac{\Lambda}{3}r^2 \ ,
\end{equation}
with the constraint $c_0^2=\frac{p^2}{q^2}$, and being $m$ in Eq. \eqref{f2} an integration constant. This solution represents a hairy black hole
(with hair parameters $p$ and $q$) with flat horizon, which is a generalization to the $SU(N)$ case of the black hole found in Ref. \cite{ACGO}.
The hair parameter is a discrete parameter (due to the $SU(N)$ structure
of the matter field in Eq. \eqref{U}), which has neither topological nor Noether charge associated
with it, and therefore represents a genuine hair. 

The event horizon of the above solution is given by 
\begin{equation} \label{rmas2}
 r_{+}=- \frac{-q^2 K\kappa a_{N}\Lambda +\left(12m\Lambda^2+\sqrt{\Lambda^3\left( (q^2K\kappa a_{N})^3 +144m^2 \Lambda \right)}\right)^{2/3}}{2\Lambda\left(12m\Lambda^2+\sqrt{\Lambda^3\left((q^2K\kappa a_{N})^3 + 144m^2 \Lambda\right)} \right)^{1/3}} \ , 
\end{equation}
where $m$ must satisfy the following condition,
\begin{equation}
m\geq \frac{q^3\left(K\kappa a_{N}\right)^{3/2}}{12\sqrt{-\Lambda}} \ ,
\end{equation}
in order to have a real value of the square root, and therefore a well defined event horizon. The above constraint reveals a minimum value
of the event horizon, which is given by 
\begin{equation}\label{rminflat}
r_{\text{min}} =  \frac{q \sqrt{K\kappa a_{N} }}{\sqrt{-\Lambda}}  \ . 
\end{equation}

\subsection{Black string}

In order to construct an extended object as solutions of the the $(3+1)$ dimensional Einstein $SU(N)$-NLSM, we will consider the following metric
\begin{equation}      \label{metric3}
 ds^2=-f(r) dt^2 +\frac{1}{f(r)} dr^2 + r^2 d\theta^2 + L^2 d\phi^2 \ ,
\end{equation}  
with $L$, for now, and arbitrary constant. On the other hand, for the matter field is sufficient to take the same $U$ field as in the
case of the flat black hole, that is 
\begin{equation} \label{F3}
 F_1(x^\mu)=0 \ , \quad F_2(x^\mu)=q \theta \ , \quad F_3(x^\mu)=p \phi \ .
 \end{equation}
Once again, with the ansatz in Eqs. \eqref{U}, \eqref{metric3} and \eqref{F3}, the complete set of Einstein $SU(N)$-NLSM equations can be solved
 for the $f$ function, obtaining
\begin{equation} \label{f3,r3}
 f(r)= -m-\frac{ q^2 K\kappa a_{N}  }{4}\log r-\frac{\Lambda}{2}r^2 \ ,  
\end{equation}
with $m$ an integration constant. Additionally, in this case, a constraint emerges from the Einstein equations that fixes the integration constant $L$
\begin{equation*}
 L^2=\frac{p^2K\kappa}{4(-\Lambda)}a_{N} \ . 
\end{equation*}
The above configuration corresponds to a black string with a compactified direction of length $L$, which is fixed in terms of the 
couplings of the theory. It is interesting to note that the black hole in the transverse section of the string is, in fact, analogous
to the charged Bañados-Teitelboim-Zanelli black hole 
\cite{Martinez}, where the NLSM coupling constant $K$ plays the role of the electric charge, and modulated for a factor that depends on the 
flavor number, namely $a_N$. We can see that, as more flavors are considered in the theory, the compactification radius of the string becomes larger.
Note also that, in order to have a well defined compactification radius, the cosmological constant $\Lambda$
must be negative. 

The event horizon of this solution is localized at
\begin{equation} \label{rmas3}
 R_{+} = \pm \frac{q}{2} \sqrt{\frac{K\kappa a_{N}}{\Lambda} \mathcal{W}\biggl[ \frac{4}{q^2 K\kappa a_{N} \Lambda}\text{Exp}(\frac{-8m}{q^2 K\kappa a_{N}  \Lambda }) \biggl]  }   \ , 
\end{equation}
where $\mathcal{W}$ denotes the Lambert-$\mathcal{W}$ function (also known as the ProductLog
function).
Although there exist a good number of solutions describing extended objects in theories of gravity 
(see \cite{Canfora:2021ttl}, \cite{BS5}, \cite{Nakas}, \cite{Sheykhi}, \cite{Estrada} for some recent results in this area),
building solutions of this kind in four dimensions in not an easy task,
and usually requires the introduction of exotic matter fields \cite{Kaloper}, \cite{Dimakis}, \cite{Muniz}.
Therefore, it is important to highlight that the black string constructed
here comes from the coupling between two very relevant physical theories, both from a theoretical and phenomenological point of view.


\section{Thermodynamics and stability}


In this section we will develop the thermodynamic analysis of the spherical black hole, the flat black hole and the black string solution constructed in the 
previous section. Also we briefly discuss the stability of these configurations and the role that the flavor number plays.

\subsection{Mass, temperature and entropy}

As can be seen from Eq. \eqref{f1}, the spherical black hole possesses an angular defect. Therefore, the thermodynamical quantities asociated to this configuation 
must be calculated using the extended thermodynamic formalism \cite{Appels}, \cite{Nucamendi}. Here we will follow the computations performed in Ref. \cite{Termo}, 
where the authors have studied the thermodynamics of the black hole in Ref. \cite{CanforaMaeda}, which correponds to the $SU(2)$
case of the spherical black hole constructed here, 
but including the Skyrme term\footnote{In fact, one can compute the mass, temperature and entropy of our spherical black hole
simply rescaling, $K \rightarrow a_N K$, and doing, $\lambda \rightarrow 0$, in Ref. \cite{Termo}.} (see also \cite{Khan}). 

The first step in this formalism is to remove the angular defect from the function $f$ in Eq. \eqref{f1}, what can be accomplished
performing the following coordinate transformations   
\begin{equation}\label{eq:ctBH}
 r = \bar{r}(1-K\kappa a_{N})^{\frac{1}{2}} \ , \qquad t = \bar{t}(1-K\kappa a_{N})^{-\frac{1}{2}} \ .
\end{equation}
By doing this change, the metric in Eqs. \eqref{metric1} and \eqref{f1}
takes the form 
\begin{equation}
 ds^2 = -F d\bar{t}^2 + \frac{1}{F} d\bar{r}^2 + (1-K\kappa a_{N})\bar{r}^2(d\theta^2+\sin^2\theta d \phi^2) \ ,
\end{equation}
where
\begin{equation*}
 F(\bar{r})=1-\frac{2\bar{m}}{\bar{r}}-\frac{\Lambda}{3}\bar{r}^2 \ , 
\end{equation*}
and $\bar{m}=m/(1-K\kappa a_{N})^{\frac{3}{2}}$. In the new coordinates, the event horizon is localized at
\begin{equation} \label{rbar}
\bar{r}_{+} = -\frac{ \Lambda +(3\bar{m}\Lambda^2+\sqrt{\Lambda^3(9\bar{m}^2\Lambda-1)} )^{\frac{2}{3}}}{\Lambda(3\bar{m}\Lambda^2+\sqrt{\Lambda^3(9\bar{m}^2\Lambda-1)} )^{\frac{1}{3}}} \ .
\end{equation}
Then, following Ref. \cite{Termo}, it is a straightforward calculation -interpreting the cosmological
constant as a bulk pressure and using the standard counterterm method to obtain a finite Euclidean action- 
to obtain the mass, the entropy and the temperature of the spherical black hole solution
in terms of the radius of its event horizon in Eq. \eqref{rbar}. We get
\begin{align}\label{eq:Ssph}
 S_{\text{sBH}} =& \pi (1-K\kappa a_{N}) \bar{r}_{+}^2 \ , \\\label{eq:Tsph}
 T_{\text{sBH}} =& \frac{1}{4\pi \bar{r}_{+}}-\frac{\Lambda \bar{r}_{+}}{4\pi} \ , \\\label{eq:Masssph}
 M_{\text{sBH}} =& (1-K\kappa a_{N})\bar{m}= \frac{1}{6}(1-K\kappa a_{N})(3-\Lambda \bar{r}_{+}^2)\bar{r}_{+} \ .
\end{align}
On the other hand, as has been showed in Refs. \cite{ACGO} and \cite{ACLV}, the thermodynamic quantities of the flat black hole and
the black string can be calculated directly using the standard formulas
\begin{equation}
 T = \frac{f'(r_+)}{4\pi} \ , \qquad S= \frac{A}{4} \ , 
\end{equation}
(where $A$ denotes the area of the event horizon) together with the ADM mass,\footnote{The same results can be derived using the phase space formalism or the counterterms methods; see
\cite{ACGO}, \cite{Kim} and \cite{Lifshitz}.} according to Ref. \cite{Vanzo}.
It follows that, for the flat black hole solution we obtain
\begin{align}\label{eq:Sflat}
 S_{\text{fBH}} =& \frac{\pi^2 p\: r_{+}^2}{2q} \ , \\ \label{eq:Tflat}
 T_{\text{fBH}}=& -\frac{K\kappa a_{N} q^2 }{16 \pi r_{+}} - \frac{\Lambda r_{+}}{4\pi} \ , \\ \label{eq:Massflat}
 M_{\text{fBH}}=& \frac{p\pi}{4q}m =  -\frac{p\pi r_{+}}{48q}\left(3 K\kappa a_{N} q^2  + 4 \Lambda r_{+}^2\right)    \ . 
\end{align}
While for the black string solution we found
\begin{align}\label{eq:Sbs}
 S_{\text{BS}} =& \frac{\pi^2 p }{4} \sqrt{-\frac{ K \kappa a_{N} }{\Lambda}}R_{+} \ , \\\label{eq:Tbs}
 T_{\text{BS}} =& -\frac{K \kappa  a_{N} q^2 }{16 \pi R_{+}} -\frac{\Lambda R_{+}}{4\pi}  \ , \\\label{eq:Massbs}
 M_{\text{BS}} =&  \frac{p\pi}{16}\sqrt{-\frac{K\kappa a_N}{\Lambda}} m = \frac{\pi R}{8} m = -\frac{p \pi}{64} \sqrt{-\frac{K\kappa a_{N}}{\Lambda}}(2\Lambda R_{+}^2 + K\kappa a_{N} q^2  \log R_{+}) \ . 
\end{align}
As expected, for the three analytic solutions, the first law of black hole thermodynamics is satisfied. 
The flat black hole and the black string solutions fulfill
\begin{equation*}
 \delta M \ = \ T \delta S \ ,
\end{equation*}
while for the spherical black hole we have 
\begin{equation*}
\delta M = T \delta S + V \delta P \ , 
\end{equation*}
where the volume $V$ and the pressure $P$ are given respectively by
\begin{equation*}
V= \frac{4}{3}\pi \bar{r}_+^3 (1-K \kappa a_N) \ , \qquad P= - \frac{\Lambda}{\kappa} \ . 
\end{equation*}

\subsection{Thermal comparison and the role of $N$}

In order to show the behavior of the thermodynamical quantities and perform a comparison between our solutions, we will set the parameters and the coupling constants as follows: Since the black string has negative cosmologial constant we set $\Lambda=-1$. Also, to have a well defined coordinate transformations in Eq. \eqref{eq:ctBH} we set $K=1$ and $\kappa=1/40$ which allows us to consider a sufficient set of $N$ values, that is $N=2,...,6$. Finally, for simplicity, we set $p=q=1$.

\begin{figure}[h!]
\centering
\includegraphics[scale=.45]{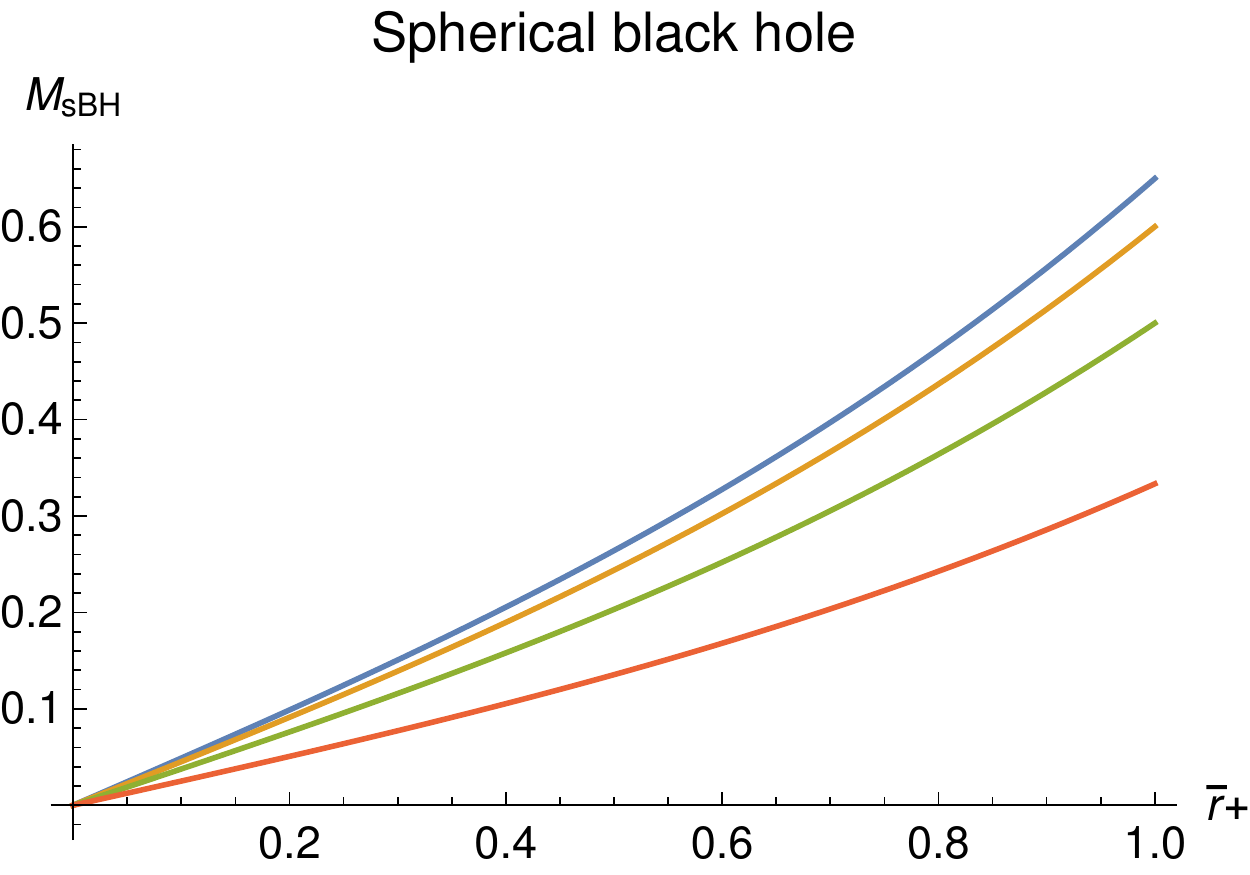}
\includegraphics[scale=.45]{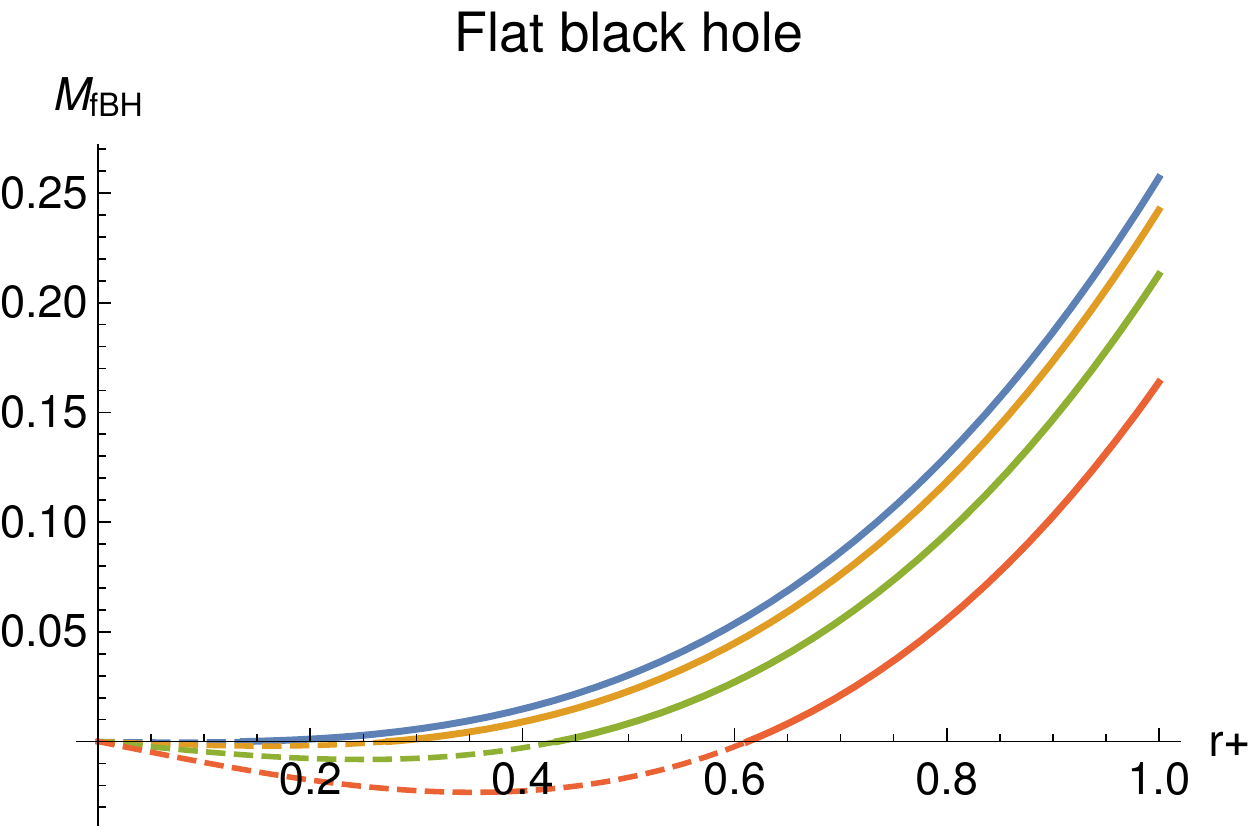}
\includegraphics[scale=.6]{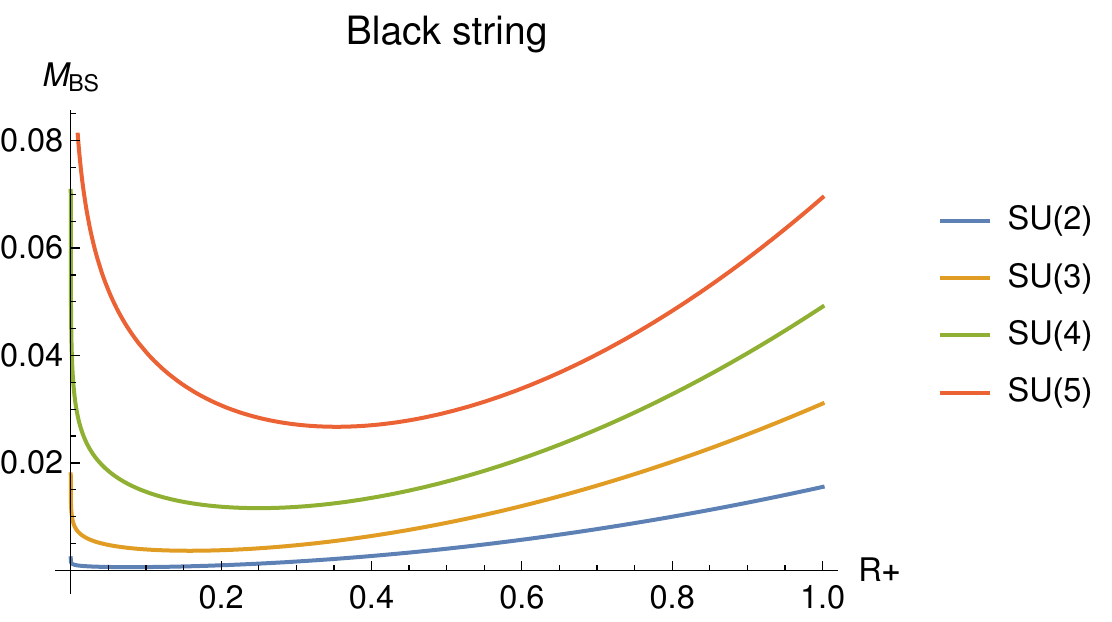}
  \caption{Mass $M$ for each solution as a function of their respective event horizon for different values of the flavor number. 
  Left: For the spherical black hole we always have a positive mass. For a fixed horizon radius the mass decreases with $N$. Center: For the flat black hole, the dashed lines highlights the sectors where the mass has negative values, for $r_+$ below the bound defined in Eq. \eqref{eq:Minrflat}. This minimum radius $r_{+\text{Min}}$ allowing positive mass grows with $N$. The mass decreases as $N$ increases. Right: For the black string, $M$ is always positive, diverges at the origin and grows quadratically at large $R_+$. It has a minimun value at $R_+=\frac{q}{2}\sqrt{\frac{K \kappa a_N}{- \Lambda}}$. For a fixed value of $R_+$, the mass of the black string increases as we increase $N$.
  }\label{fig:Masas}
\end{figure}

The black holes and black string constructed in Sec. 3 have different physical properties in which the value of $N$ plays a very important role.
For instance, the mass of our solutions, according to the Eqs. \eqref{eq:Masssph}, \eqref{eq:Massflat}, \eqref{eq:Massbs} depend explicitly on $N$. These dependence is shown in Fig. \ref{fig:Masas}, where we have plotted the mass of the solutions as function of their proper event horizon for several values of $N$. We can see that the three configurations present different features. First, for the spherical black hole we see that its mass is a monotonically increasing and always positive function of its event horizon. 
On the other hand, the flat black hole has a lower bound for its event horizon, setting the minimun radius with a non-negative mass that depends on $N$, and it can be obtained explicitly
from Eq. \eqref{eq:Massflat}, being
\begin{equation}\label{eq:Minrflat}
r_{+}\geq \frac{q}{2}\sqrt{\frac{3 K\kappa a_N}{-\Lambda}}=r_{+\text{Min}} \ .
\end{equation}
Here the mass also grows monotonically with the radius. For the black string, the mass is positive for any horizon radius, but, in contrast with the spherical black hole case, the black string mass is no monotonic; as $R_+$ decreses the mass increases, and we have a minimum value of the mass located at $R_+=\frac{q}{2}\sqrt{\frac{K \kappa a_N}{- \Lambda}}$. Past this critical value the mass starts to grow as $R_+$ grows.

Regarding the $N$ dependence, we can see two different behaviors. For a fixed value of the event horizon, both black hole masses decrease as we increase the value of $N$. In contrast with what happens for the black hole solutions, for a fixed value of horizon the mass of the black string increases as we increase the flavor number of the configuration.

\begin{figure}[h!]
  \centering
    \includegraphics[scale=.5]{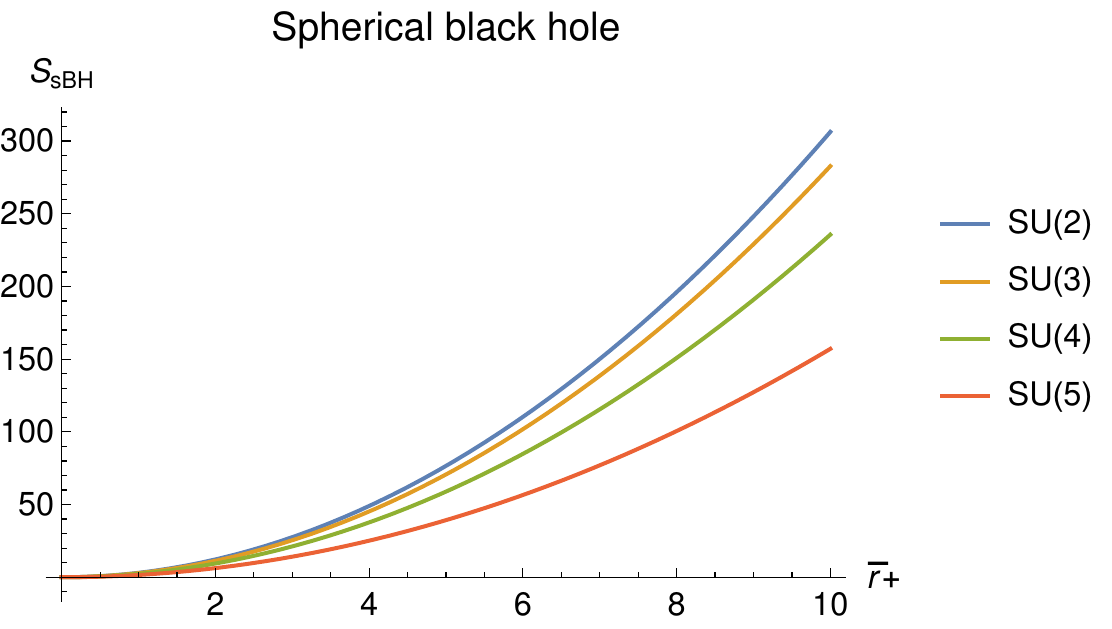}
    \includegraphics[scale=.5]{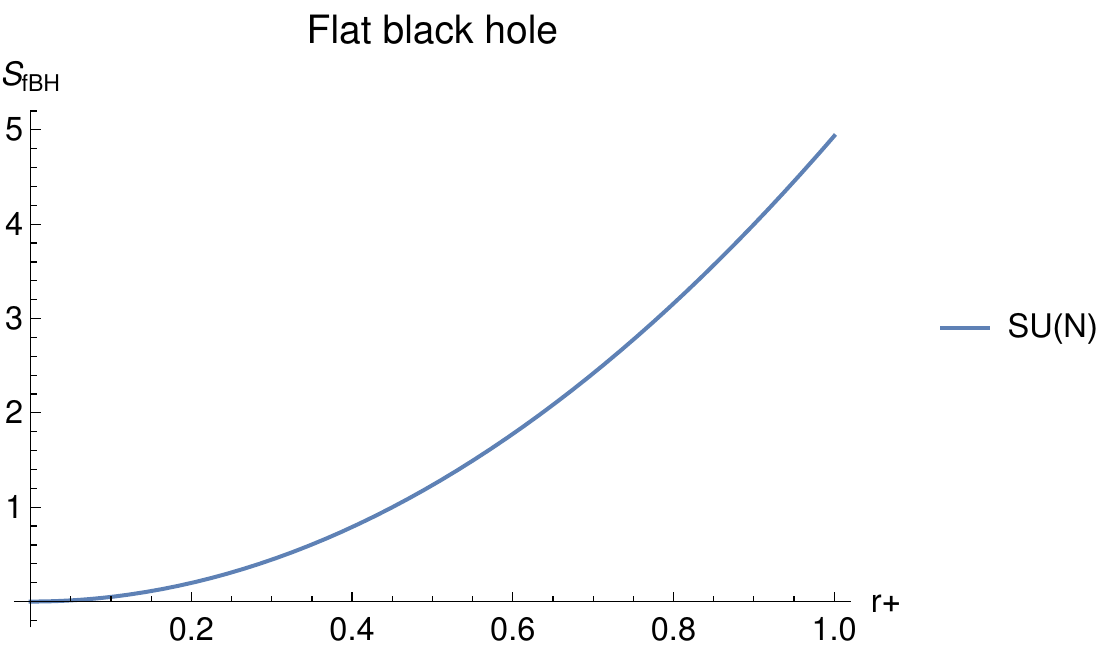}
    \includegraphics[scale=.5]{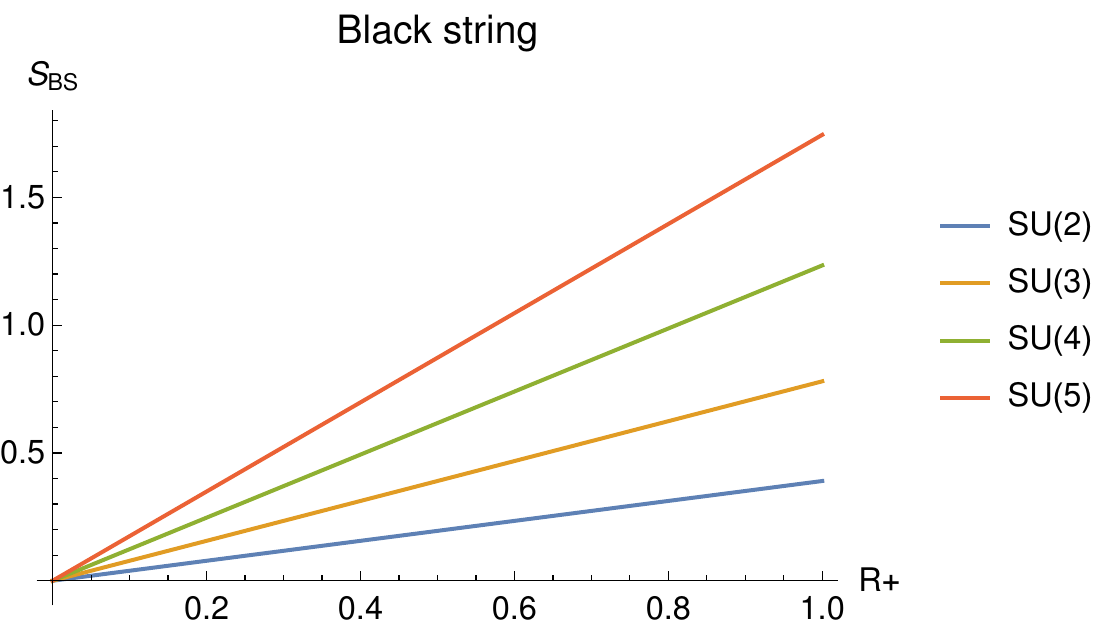}
  \caption{Entropy $S$ for each configuration as a function of their respective event horizon for different values of the flavor number. Left: The spherical black hole entropy behaves as $ \bar{r}_+^2$. For a fixed horizon radius $S$ decreases as $N$ grows. Center: The flat black hole entropy behaves as $r_+^2$ and it is independent of the group dimension defined by $N$. Right: The entropy of the black string. Here, $S \sim R_+$, and for a fixed radius $S$ grows with the flavor number.
  }\label{fig:Entropias}
\end{figure}

In Fig. \ref{fig:Entropias} we see the entropy of each configuration as a function of their respective event horizon and their particular $N$ dependence (the entropy comparison between these configurations will be given below). Remarkably, all three configurations presents very different features regarding the value of the flavor number. In fact, the entropy of the spherical black hole behaves as $\bar{r}_+^2$, according to Eq. \eqref{eq:Ssph}. We see that for a fixed radius, the entropy decreases as we increase $N$. For the flat black hole, according to Eq. \eqref{eq:Sflat}, we get the same cuadratic behavior for its entropy in terms of $r_+$, but now $S$ does not depend on $N$. For the black string entropy, given in Eq. \eqref{eq:Sbs}, the behavior in terms of $R_+$ is linear and its dependence of $N$ is opposite to that of the spherical black hole, that is, for a given event horizon, the entropy of the black string grows as we increase $N$.
The above suggests that, for the spherical black hole, the thermally favored configurations are those with the lowest number of flavors, while for the black string they are those with the highest number of flavors. It is worth mentioning that in Ref. \cite{ACGO} was shown that, in the $SU(2)$ case, the flat black hole is always
thermodynamically favored with respect to the corresponding black hole with vanishing pionic field, now, as we can see, this fact remains unchanged for arbitray $N$.

\begin{figure}[h!]
  \centering 
   \includegraphics[scale=.5]{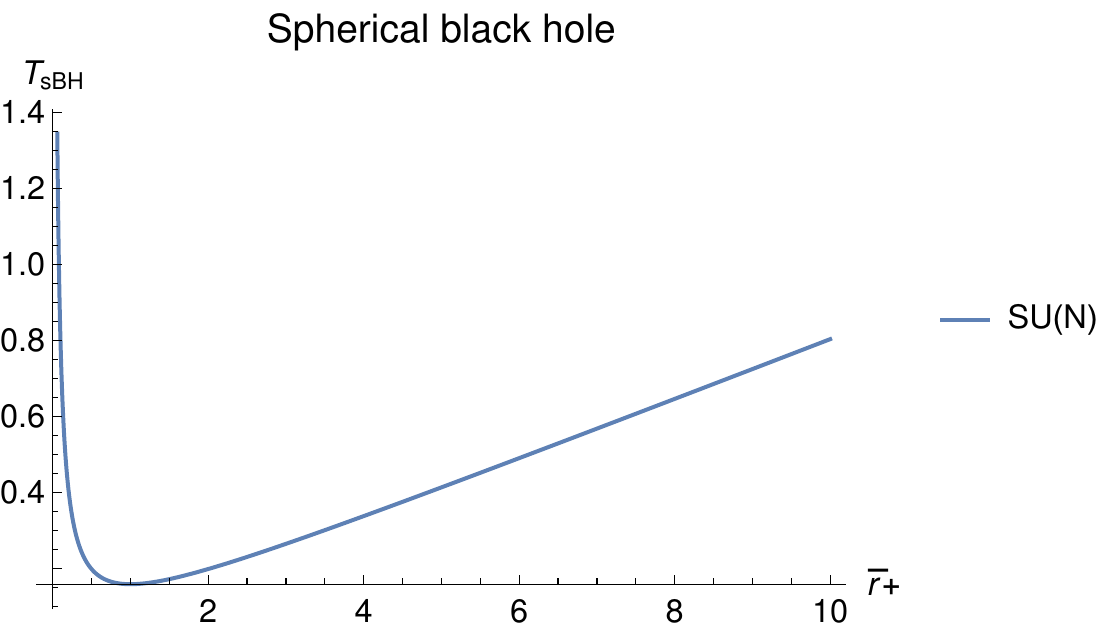}
    \includegraphics[scale=.5]{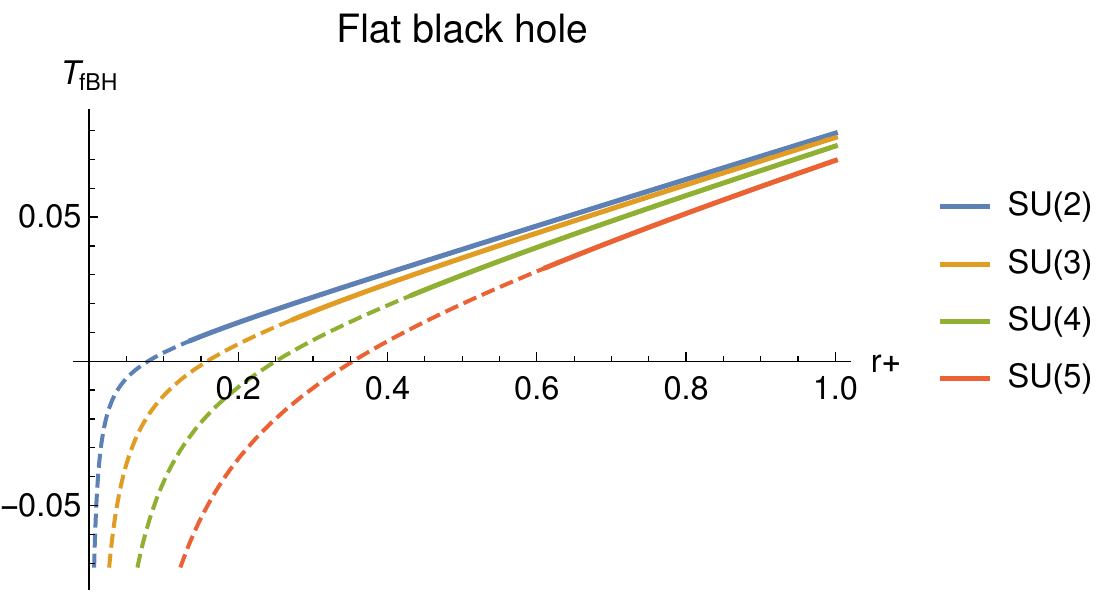}
    \includegraphics[scale=.5]{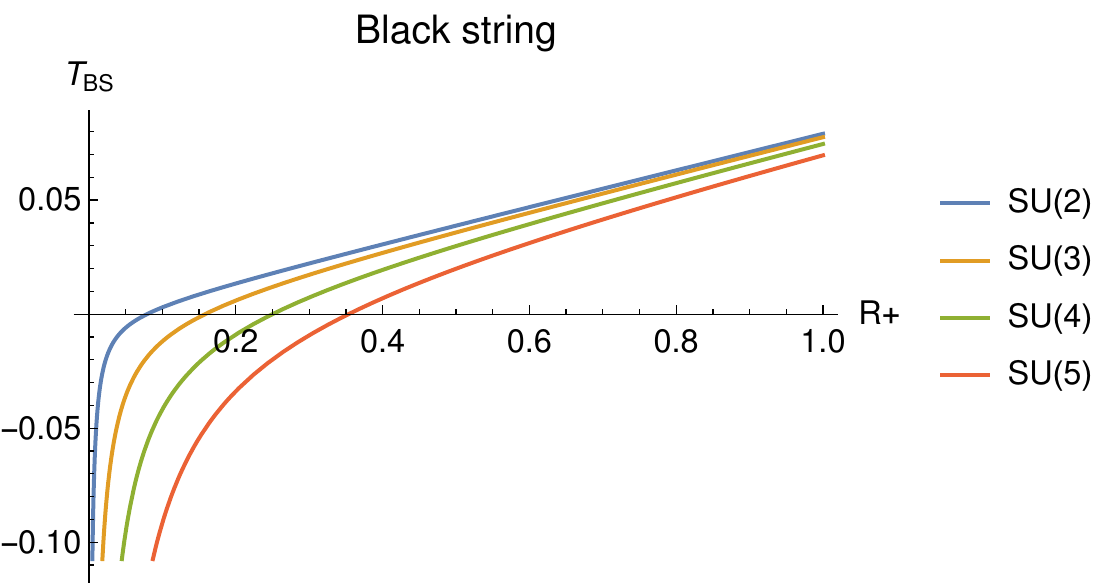}
  \caption{Temperature $T$ as a function of their respective event horizon for different values of $N$.
  Left: The spherical black hole temperature is independent of $N$, always positive and has a minimum where $T(\bar{r}_{\text{min}}=1)=1/(2\pi)$.
  Center: Temperature for the flat black hole. The dashed lines represent the temperature for radius below the bound defined in Eq. \eqref{eq:Minrflat}. We see that $T$ goes to $-\infty$ as $r_+\rightarrow 0$, but these values are not allowed due to the restrictions given by the positivity of the mass. For $r>r_{+\text{Min}}$ we have positive temperatures. For a fixed $r_+$, $T$ decreases as we increase $N$.
  Right: Black string temperature. Here $T$ goes to $-\infty$ as $R_+\rightarrow 0$. At a certain horizon radius value that depends on $N$, the temperature becomes positive. For a fixed $R_+$, $T$ decreses with the increase of $N$.
  }\label{fig:Tes}
\end{figure}

The temperature of the solutions as a function of their event horizon for different values of $N$ are given in Fig. \ref{fig:Tes}. 
The spherical black hole temperature, according to Eq. \eqref{eq:Tsph}, is independent of the flavor number and has the same behavior for all the Lie groups here considered. It is always positive and has a minimum where $T(\bar{r}_{\text{min}}=1)=1/(2\pi)$.

The temperature for the flat black hole, according to Eq. \eqref{eq:Tflat}, is a monotonically increasing function of $r_+$. The dashed lines represent the temperature for radius below the bound defined in Eq. \eqref{eq:Minrflat}. We see that $T$ goes to $-\infty$ as $r_+\rightarrow 0$, but these values are not allowed due to the restrictions given by the positivity of the mass. For $r>r_{+\text{Min}}$ we have positive temperatures. For a fixed $r_+$ we have that $T$ decreases as we increase $N$. Finally, for the temperature of the black string, according to Eq. \eqref{eq:Tbs}, we have the same functional expression that in the flat black hole case, but here there is no restriction for a minimum radius, then, we have allowed black string solutions with small enough horizon radius having negative temperature. In fact, $T$ goes to $-\infty$ as $R_+\rightarrow 0$, then at a certain horizon radius, depending on $N$, $T$ becomes positive. Also, for a fixed $R_+$ the temperature of the black string decreses with the increase of $N$.

On the other hand, as it is well known, black strings in general relativity suffer from the Gregory-Laflamme instability
\cite{GregoryLaflamme1}, \cite{GregoryLaflamme2}.
Although this instability has a perturbative nature, it is also manifested in the comparison of the entropies
between the black hole and the black string. When the entropies are compared as a function of their masses,
there is a critical mass below which the black hole has higher entropy, and it is therefore thermally favored. 
Above such critical mass the behavior is the inverse. 
Here, to study the possible existence of a Gregory-Laflamme transition we will take the latter path because the study of the perturbative stability
involves linearizing and decoupling a set of very complicated equations.\footnote{We hope to return to the study of the perturbative stability in a future publication.}
In fact, using the equations of the previous subsection, we can compare the entropies as functions of the radius at fixed equal mass. 

\begin{figure}[h!]
  \centering
   \includegraphics[scale=.7]{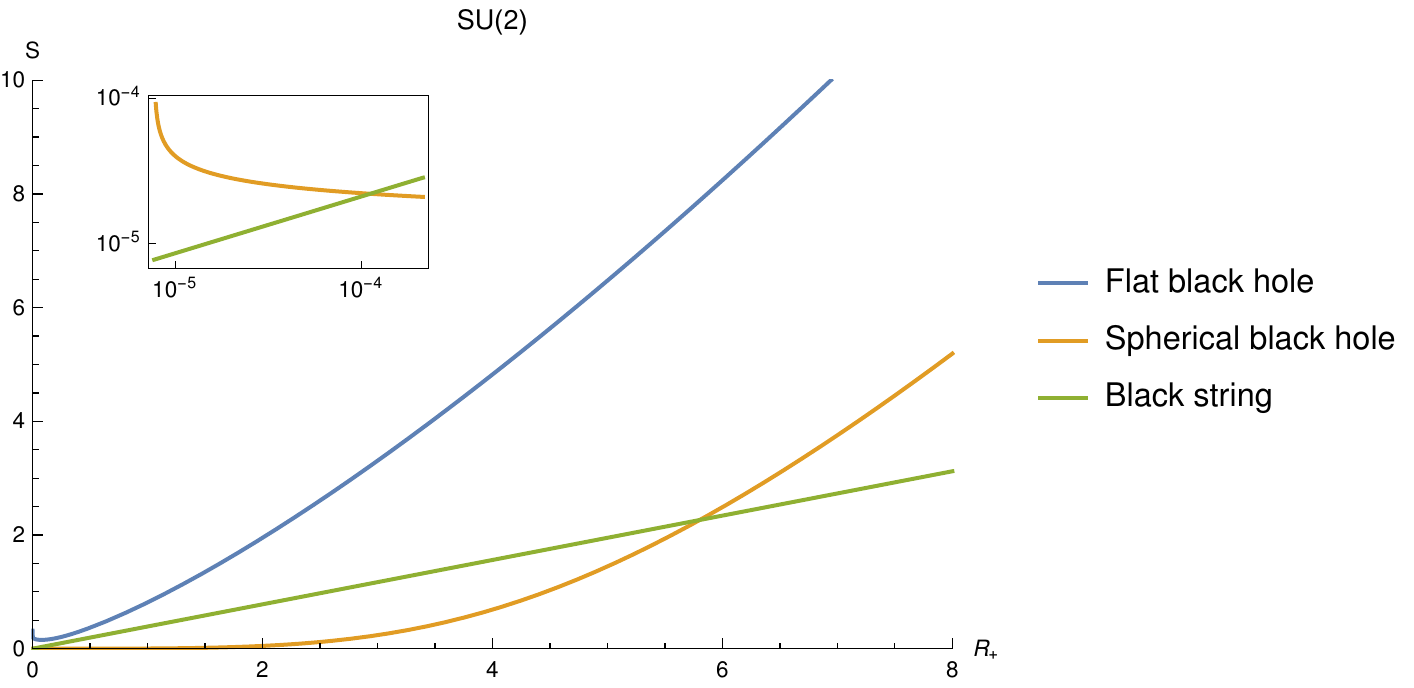}\\
    \includegraphics[scale=.7]{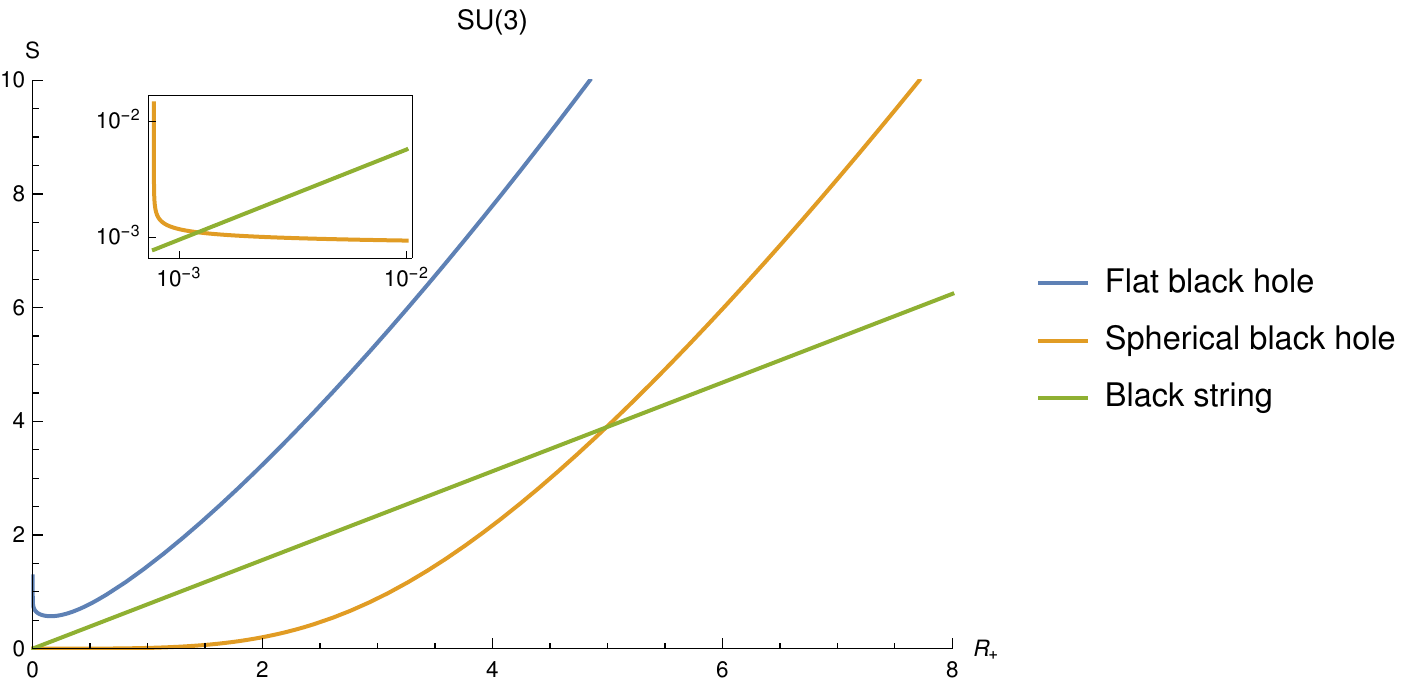} 
  \caption{Comparison of the entropies of the spherical black hole, the flat black hole and the black string, in terms of the black string event horizon at equal mass. 
   Up: The $SU(2)$ case. The entropy of the flat black hole is always larger. There are transitions between the spherical black hole and the black string. Inset: for small values of $R_+$ the entropy of the black hole is greater than the entropy of the black string. During a finite $R_+$ interval the black string has greater entropy than the spherical black hole. For large enough $R_+$ a new transition occurs, getting the same behavior as for small radii.
    Down: The $SU(3)$ case. As we increase $N$ the interval where the transitions between the spherical black hole and black string occurs becomes smaller.
  }\label{fig:SNs}
\end{figure}

Explicitly, we will write the entropies of the spherical and flat black holes, in Eqs. \eqref{eq:Ssph} and \eqref{eq:Sflat}, in terms of the event horizon of the black string, $R_+$, by demanding equal masses defined in Eqs. \eqref{eq:Masssph}, \eqref{eq:Massflat} and \eqref{eq:Massbs}, that is
\begin{equation}
M_{\text{sBH}}= M_{\text{fBH}}=M_{\text{BS}}\ .
\end{equation}
From the above condition we can solve $\bar{r}_{+}$ and $r_+$ in terms of the black string radius and then, from Eqs. \eqref{eq:Ssph} and \eqref{eq:Sflat}, have all the entropies as functions of $R_+$. The analytical expressions for the $S(R_+)$ obtained in this way are very cumbersome, however the comparison of the entropies in Fig. \ref{fig:SNs} will clear things out. 

In Fig. \ref{fig:SNs} is shown that the flat black hole solution has a higher entropy than the black string and spherical black hole solutions at any value of $R_+$. Also, we can see the existence of two transitions occurring between the spherical black hole and the black string. For small values of $R_+$ we get that the entropy of the black hole is greater than the entropy of the black string, as we see in the inset. Between these two critical points we have the opposite behavior and the black string entropy is larger than the black hole entropy. For large enough $R_+$ we have again another transition between these two solutions, obtaining the same behavior as for small radii. Finally, as $N$ increases, the interval defined by these two critical points becomes smaller, as we see explicitly in the comparison with the $SU(3)$ case. Also, it is worth to notice that for $N>2$ we have the same behavior reported in Ref. \cite{ACLV} for the entropies of the flat black hole and black string solutions.

\begin{figure}[h!]
  \centering
   \includegraphics[scale=.5]{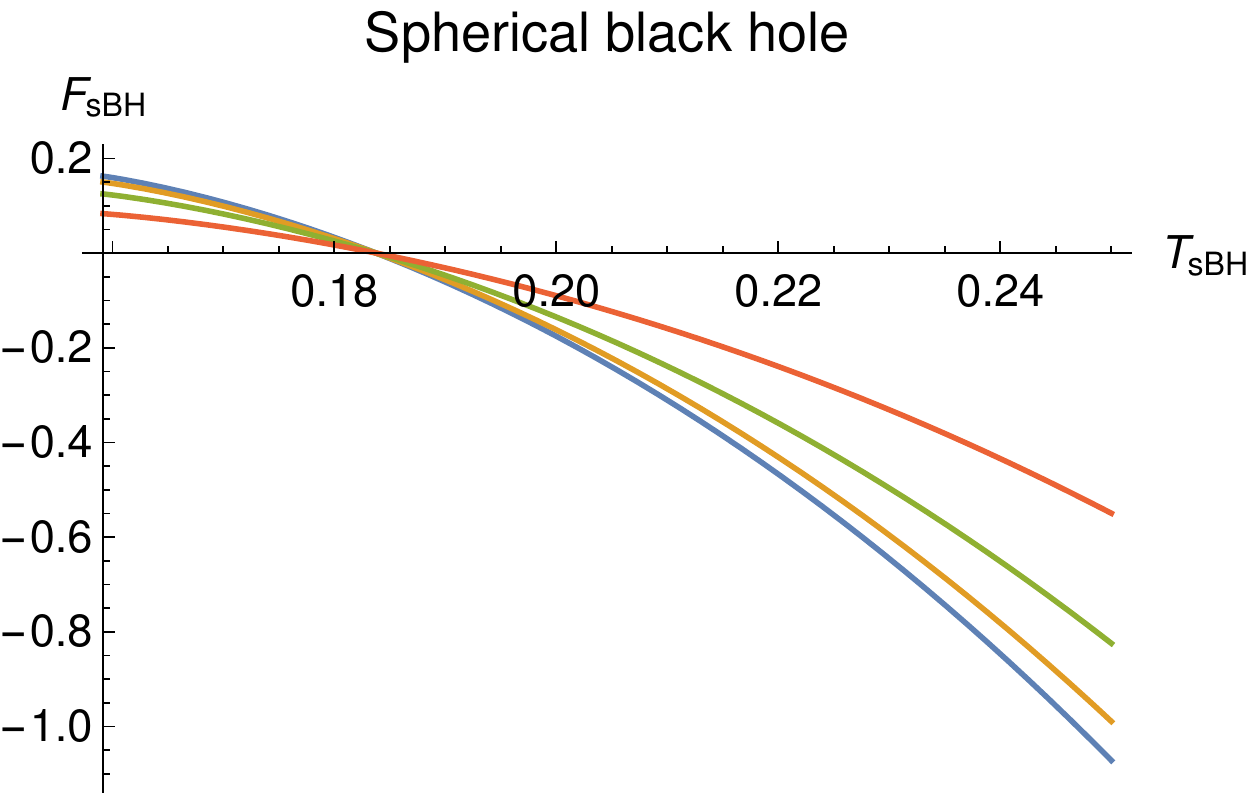}
    \includegraphics[scale=.5]{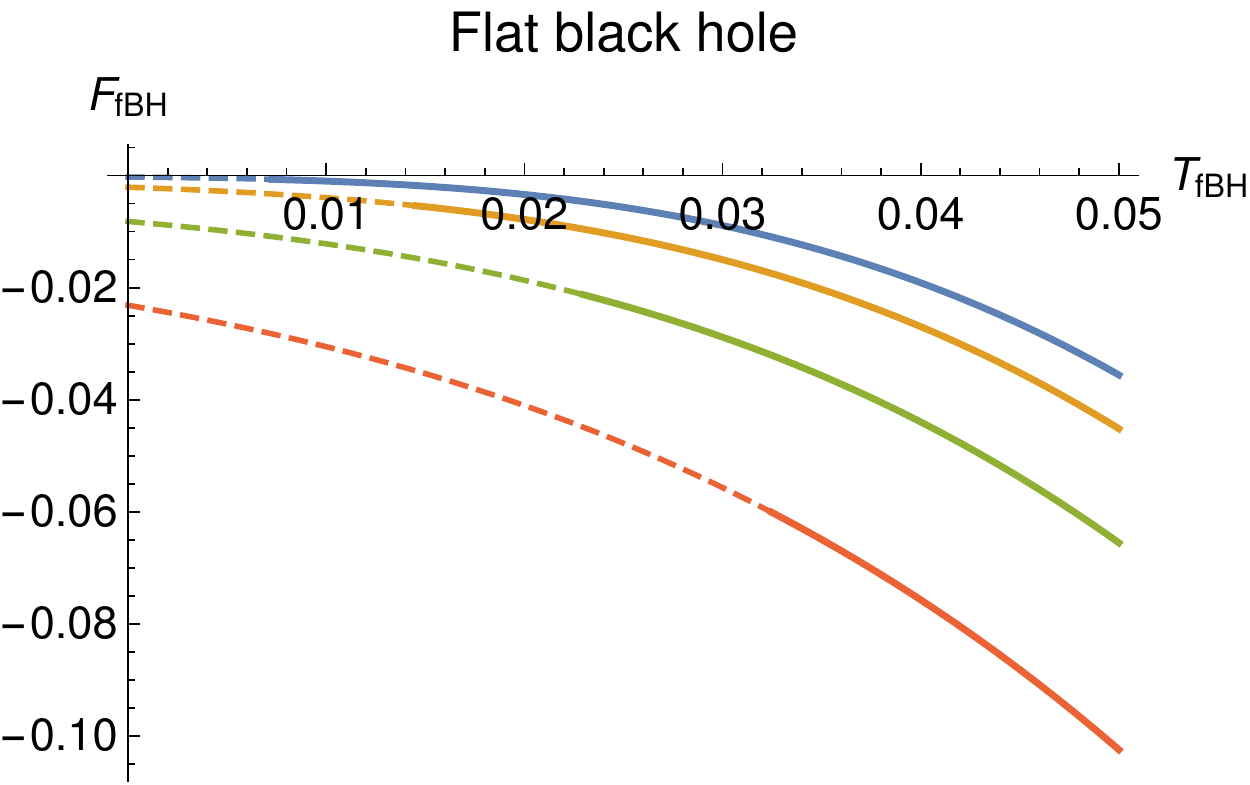}\\
    \includegraphics[scale=.7]{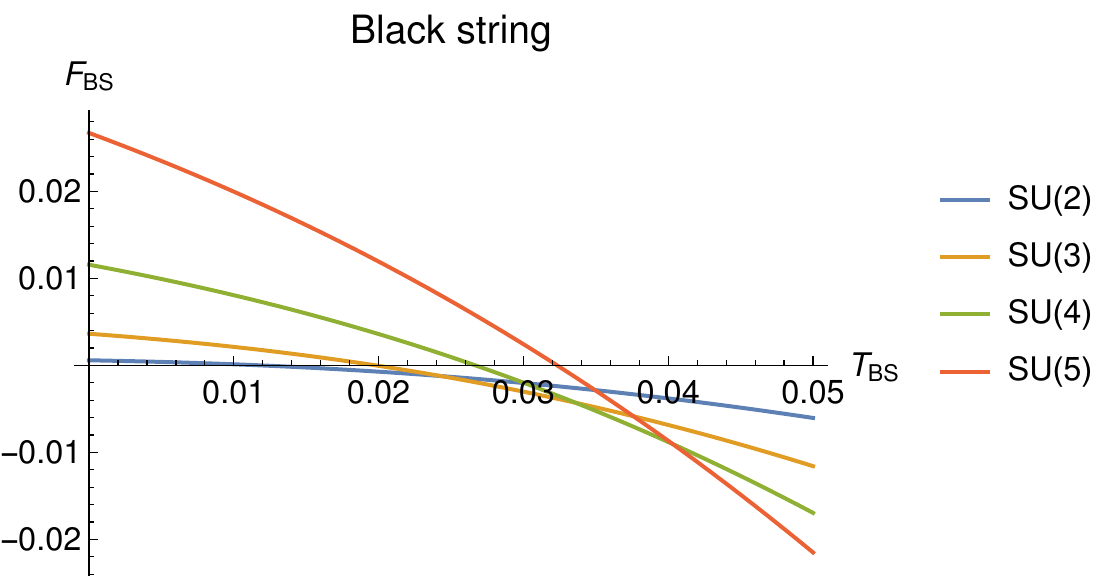}
  \caption{Free energy $F$ as a function of $T$ for different values of $N$.
   Up-Left:
    The spherical black hole case. For small and fixed values of $T$, $F$ is lower as we increase $N$. By increasing the temperature transitions starts to happen, and for high enough values of $T$ the lower free energy is for the case with $N=2$.
   Up-Right:
      The flat black hole case. The favored configuration is always the one with the higher flavor number. The dashed lines represent the sectors with non-positive mass.
   Down:
     The black string case. For small and fixed values of $T$ the configuration with $N=2$ has less free energy. As we increase the temperature transitions starts to happen, and for high enough values of $T$ we get that the free energy is lower as we increase $N$.
  }\label{fig:FTs}
\end{figure}

Finally, we compute the free energy, $F=M-TS$. For the spherical black hole, the flat black hole and the black string, the free energy given in terms of their event horizons, is respectively given by
\begin{eqnarray}\label{eq:Fsph}
F_{\text{sBH}}&=&-\frac{1}{12}\left(K\kappa a_N -1\right)\left(\bar{r}_{+}^2\Lambda +3 \right) \bar{r}_{+} \ , \\\label{eq:Fflat}
F_{\text{fBH}}&=&-\frac{1}{32}\frac{\pi p r_{+}}{q}\left(K\kappa a_{N}q^2 -\frac{4}{3}\Lambda r_{+}^2 \right) \ , \\\label{eq:Fsb}
F_{\text{BS}}&=&-\frac{1}{64}\pi p \sqrt{-\frac{K\kappa a_{N}}{\Lambda}}\left(K\kappa a_{N}q^2 \left(\log R_{+}-1\right) -2\Lambda R_{+} ^2\right) \ .
\end{eqnarray}
In order to obtain the free energy in terms of the temperature is sufficient to invert Eqs. \eqref{eq:Tsph}, \eqref{eq:Tflat} and \eqref{eq:Tbs}, take the only positive root and substitute respectively into Eqs. \eqref{eq:Fsph}, \eqref{eq:Fflat} and \eqref{eq:Fsb}.
The resulting analytical expressions of the free energy as a function of the temperature are very complicated, but they are not needed for our purposes. Instead, in Fig. \ref{fig:FTs} we graph $F(T)$ for different values of $N$. First, in the spherical black hole case, for low but fixed values of $T$, we see that the free energy decreases for higher values of $N$. As we increase the temperature transitions starts to happen, and for high enough values of $T$ we get that the lower free energy is for the configuration with $N=2$. In the flat black hole case, the favored configuration is always the one with the higher flavor number. The dashed lines in the plot represent the sectors with non-positive mass. Finally, in the black string case, for low values of $T$, the configuration with $N=2$ has the lowest free energy. As we increase the temperature transitions start to happen, and for high enough values of the temperature we get that the free energy is lower for the configuration with the highest $N$.

\subsection{A comment on classical stability}

As we have mentioned before, the study of the classical stability of the solutions constructed above is 
a very difficult problem. For gravitating solutions coupled to matter fields (and, in particular, with non-Abelian fields) 
one must deal with a big set of coupled diferential equations. In our case, an extra complexity is that each solution has different symmetries and,
therefore, the perturbation to be considered must be (in principle) different for each case. For example, in
the case of the black string, the ``most dangerous” perturbation (that is, the perturbation that respects the
symmetries of the system and can generate unstable modes) must include tensor, vector and scalar modes, in addition
to the corresponding perturbations to the three degrees of freedom of the NLSM matter field, which makes
the system very difficult to decouple. \footnote{In fact, the study of the stability of extended objects without matter fields is itself a
complex problem, as can be seen in the pioneer work of Gregory and Laflamme \cite{GregoryLaflamme1}.}
Even more, the generalization to the $SU(N)$ case add different factors to deal with in the linearized equations for each group.
Something similar happens with the black hole solutions.

However, although a general study of the classical stability requires a lot of work,\footnote{A more detailed study of the classical stability
of our solutions will appear in a future publication.} we can mention for now that there are two relatively simple paths
that provide signals about the behavior of these solutions under small perturbations.

The first path is computing the heat capacity. It is known that, for extended objets, the computation of the heat
capacity 
\begin{equation*}
 C \ = \  T\left(\frac{\partial S}{\partial T} \right) \ , 
\end{equation*}
is directly related to the classical stability. In fact, the correlated stability conjecture claims that
gravitational systems with translational symmetry and infinite extent exhibit a Gregory-Laflamme instability if and only if they have a local
thermodynamic instability \cite{Gubser}, \cite{Harmark}. Then, we need to study  the sign of the heat capacity of the solutions.

For the spherical black hole, the flat black hole and the black string we found 
\begin{align}
 C_{\text{sBH}} \ =& \ \frac{\pi\left(1-K\kappa a_{N}\right)}{2 \bar{r}_{+}^3\Lambda^2}\left(1-\Lambda \bar{r}_{+}^2 \right)\left(1-\bar{r}_{+}^2\Lambda+\bar{r}_{+}\sqrt{(1+\bar{r}_{+}^2\Lambda)^2/\bar{r}_{+}^2} \right)^2 /\sqrt{(1+\bar{r}_{+}^2\Lambda)^2/\bar{r}_{+}^2} \ , \\
 C_{\text{fBH}}\ =& \ \frac{\pi^2 p r_{+}^2}{q}\frac{\left(4 \Lambda r_{+}^2  + q^2 K \kappa  a_{N}\right)}{\left( 4\Lambda r_{+}^2  -q^2 K \kappa   a_{N}\right) } \ , \\
 C_{\text{BS}} \ =& \ \frac{\pi^2 p R_{+} \sqrt{ K\kappa a_{N}}}{4\sqrt{-\Lambda}}\left(\frac{4\Lambda R_{+}^2 +q^2 K \kappa a_{N}}{4\Lambda R_{+}^2-q^2 K \kappa a_{N} }\right) \label{Cbs} \ .
\end{align}
We can see that, in order to have a positive value of $C$, the following constraints must be satisfied
\begin{equation}  \label{constraintC}
\bar{r}_{+}< 0 \ , \qquad r_{+}> \frac{q}{2}\frac{\sqrt{K \kappa a_{N}}}{\sqrt{-\Lambda}} \ , \qquad R_{+}>\frac{q\sqrt{K\kappa a_{N}}}{2\sqrt{-\Lambda}} \ . 
\end{equation}
Combining the above constraints with the minimum radii of the event horizon of the configurations in Eqs. \eqref{rminsph}, \eqref{rminflat} and \eqref{rmas3}, we can see the following:
First, for the black string solution the heat capacity takes positive and negative values, which suggests a perturbative instability where the horizon radius is small.
This is in accordance with the Gubser-Mitra conjecture, which postulates that a thermal instability necessarily leads to a perturbative instability, and also with the thermal analysis showed in the previous subsection. Second, the spherical black hole has always a negative heat capacity, as in the case on the black hole without the coupling with pions. Finally, for the flat black hole the heat capacity is always positive due to the minimum radius of the horizon always satisfies the requirement in Eq. \eqref{constraintC}. Note that this analysis is independent on the value 
of $N$ for the black hole solutions, while for the black string the instability region increases with the flavor number.

A second approach is to consider a particular type of perturbations, namely radial perturbations, on the four degrees of freedom of the solutions \cite{Droz}, \cite{Droz2}, \cite{Shiiki};
the function $f$ in the metric and the three pionic degrees of freedom of the matter field $F_i$.
In fact, one can check that considering perturbations of the form 
\begin{equation*}
 f(r) \rightarrow f(r) + \epsilon e^{i \omega t} P_0 (r) \ , \qquad  \ F_i(r) \rightarrow F_i(r) + \epsilon  e^{i \omega t} P_i (r) \ ,  \qquad i=1,..., 3 \ , \qquad \epsilon \ll 1 \ , 
 \end{equation*}
the linearized field equations can be reduced to a single Schr\"odinger equation for one of the components of the perturbation (redefining appropriately such component), at least in the case of the spherical black hole and the black string solution. This opens the possibility of studying the existence of unstable modes in these configurations, which is part of a work in progress. For the flat black hole this construction cannot be carried out in the same way, which suggests the need to consider more general perturbations to study its stability.

Summarizing, taking into account both the thermal arguments and the partial perturbative results we can conjecture the following:
The black string and the spherical black hole solutions possess unstable modes, and therefore transitions between these configurations can occur.
The black string is expected to suffer from a Gregory-Laflamme instability which can be seen when considering perturbations with tensor, vector and scalar modes.\footnote{The study of the possible Gregory-Laflamme instability of the black string solution
under general perturbations is a work in progress.} On the other hand, the previous arguments also suggest stability of the flat black hole, although a more elaborate treatment is necessary to guarantee this.


\section{Conclusions}


In this work we have constructed black holes and a black string as analytical solutions of the Einstein $SU(N)$-NLSM theory, 
generalizing the results in Refs. \cite{CanforaMaeda}, \cite{ACGO}, \cite{ACLV} to the $SU(N)$ symmetry group case.
First, the spherical black hole is characterized by an angular defect that depends on the flavor number $N$, 
and it is asymptotically the Anti-de Sitter version of the Barriola–Vilenkin space-time.
The second solution, the black hole with flat horizon, possess discrete hair parameters, and it is asymptotically locally Anti-de Sitter.
On the other hand, the black string constructed here possess the geometry of a charged Bañados-Teitelboim-Zanelli black hole in the transverse section with the pions coupling fulfilling the role of the electric charge.
This solution exist in a space-time with negative cosmological constant and its compactification radius is fixed by the flavor number and the 
couplings of the theory. We have shown that, for fixed radius of the event horizon, the mass of the black holes decrease with $N$, while for the black string the behavior is the opposite.
We have also seen that $N$ plays a very important role in the thermodynamics of these configurations. 
Looking at the entropy, for the spherical black hole, the thermally favored configurations
are those with the lowest number of flavors, while for the black string they are those with the highest $N$. The flat black hole entropy is independent of the flavor number.
Also, comparing the entropies at equal mass, one can see that the flat black hole is the thermally favored configuration between the three solutions, and that transitions between the spherical black hole and the black string can occur.
From the computation of the free energy we have that for the spherical black hole and black string, as we increase the temperature, transitions starts to happen between configurations with different values of $N$.
It is important to mention that the black hole solutions constructed here can be generalized to solutions of the $SU(N)$ Skyrme model, which is a work in progress.

\subsection*{Acknowledgments}

The authors are grateful to Fabrizio Canfora, Gustavo Dotti, Daniel Flores-Alfonso and Julio Oliva for many enlightening comments.
M. L. is funded by FONDECYT post-doctoral Grant No. 3190873. A. V. is funded by FONDECYT post-doctoral Grant No. 3200884.
C. H. is partially funded by the National Agency for Research and Development ANID - PAI Grant No. 77190078 
and also thanks the support of FONDECYT Grant No. 1181047.

\end{document}